# How well do lattice simulations reproduce the different aspects of the geometric Schwinger model

H. Dilger and H. Joos

Deutsches Elektronen-Synchrotron DESY, Notkestr. 85, 22603 Hamburg, FR Germany

We compare continuum and lattice formulation of the geometric Schwinger Model on the torus. The lattice reproduces the $U(1)_A$ anomaly, related to non-trivial topological gauge configurations and zero modes.

## 1. THE CONTINUUM MODEL

We study non-perturbative effects related to anomalous axial symmetry breaking in the Geometric Schwinger Model (GSM) on the torus $\mathcal{T}_2$ [1–4]. 'Geometric' means formulated with Dirac-Kähler fermions. This is equivalent to a 2-flavour Schwinger model. So the action reads ($b = 1, 2$)

$$S = \int_{\mathcal{T}_2} dx [\frac{1}{2} F(x)^2 + \bar{\psi}^{(b)} \gamma^\mu (\partial_\mu - ieA_\mu)\psi^{(b)}]. \quad (1)$$

The $U(1)$ gauge field on $\mathcal{T}_2$ can be written

$$A_\mu(x) = \partial_\mu a(x) + \epsilon_\mu^{\ \nu} \partial_\nu b(x) + t_\mu + \frac{2\pi k}{eV} \epsilon_\mu^{\ \nu} x_\nu, \quad (2)$$

$\partial_\mu a(x)$ is a pure gauge, $\epsilon_\mu^{\ \nu}\partial_\nu b(x)$ is a gauge field in Lorentz gauge, $t_\mu$ is the so-called toron field, the last term is a representative of the Chern class with topological quantum number $k \in \mathbf{Z}$.

Depending on $k$ we have zero modes (ZMs) of the Dirac operator according to the index theorem. In the GSM these are $2k$ left-handed (right-handed) ZMs for $k > 0$, ($k < 0$), 0 (2 vectorlike) ZMs for $k=0$ depending on $t_\mu$ [2].

VEVs are evaluated with help of the effective action and the fermion propagators in [3]. Anomalous axial symmetry breaking appears in the correlation functions of currents and scalars. They get 3 kinds of contributions: connected (c-) and disconnected (d-) in the topologically trivial sector, and ZM contributions in the sectors $k = \pm 1$. The latter appear by an exact treatment of the ZMs in the fermion integration [5].

The isovector current (IVC) correlation function $j^i_\mu = \bar{\psi}\gamma^\mu \tau^i \psi$ has only a connected part

$$< j^k_\mu(x) j^l_\nu(0) > = \delta_{kl} \frac{-2}{\pi} \epsilon_{\mu\rho}\epsilon_{\nu\sigma} \partial_\rho \partial_\sigma G_0(x) + c. \quad (3)$$

$G_0(x)$ is the massless Greens function on $\mathcal{T}_2$ with $-\Box G_0(x) = \delta^{(\prime)}(x)$ ($^\prime$:omitted ZM).

The correlation function of the isoscalar current (ISC) $j^0_\mu = \bar{\psi}\gamma^\mu \tau^0 \psi$ is

$$< j^0_\mu(x) j^0_\nu(0) > = \frac{-2}{\pi} \epsilon_{\mu\rho}\epsilon_{\nu\sigma} \partial_\rho \partial_\sigma G_m(x). \quad (4)$$

This results as the c-contribution, which is equal to the IVC case eq. (4), is canceled by the massless part of the d-contribution. It remains the massive Greens function $G_m(x)$ of the d-part, leading to a non-vanishing divergence of the axial isoscalar current, the $U(1)_A$ anomaly.

Finally the 8 (pseudo)scalar components $s^i = \bar{\psi}\tau^i\psi$ ($s^i_5 = \bar{\psi}\gamma^5\tau^i\psi$), $i = 0, 1, 2, 3$, split into two groups $s^i_+$ and $s^i_-$. We get with the Jacobi $\Theta$-function $\theta(x) \equiv \Theta((x_1+ix_2)/L_1 | iL_2/L_1)$

$$< s^k_\pm(x) s^l_\pm(0) > = \delta_{kl} C \frac{|\theta_1(\frac{x}{2})|^4 + |\theta_3(\frac{x}{2})|^4}{|\theta_1(x)|}$$
$$\times [\exp(2\pi \overline{G}_m(x)) \pm \exp(-2\pi \overline{G}_m(x))]. \quad (5)$$

This splitting corresponds to a splitting of the $U(2)_L \times U(2)_R$ flavour octet into 2 quadruplets by the $U(1)_A$ anomaly. It is realized by the ZM contributions from the topological sectors $k = \pm 1$ (the second term in the second line of eq. (5)).

## 2. LATTICE SIMULATIONS

The lattice version of the GSM leads to staggered fermions coupled to a compact gauge field. For our MC results we used the hybrid Monte Carlo algorithm [6] on a 16×6-lattice. Let us briefly discuss its behaviour in the context of the topological features of our model [1].

It is possible to define a quantized lattice observable $Q_{top} \in \mathbf{Z}$, which coincides with the Chern index in the continuum limit. A difficulty arises from the potential barrier between the different topological sectors on the lattice, given by $Q_{top}$. Therefore it was necessary to use the so-called instanton hits [7, 8], known from pure $QED_2$, to which we added simultaneous toron shifts.

Another critical point is the measurement of contributions from the topologically non-trivial sectors corresponding to the ZM contributions in the continuum. We expect high values for the observables in question while the fermion determinant is very small in these sectors. The reason are approximate ZMs as remnants of the index theorem on the lattice [8]. Therefore we performed separate measurements in the different sectors.

We determine weak and strong coupling regions of $\bar\beta$ ($1/e^2$ in lattice units) from the behaviour of gauge field observables. Corresponding to the field strength correlation in the continuum we measure the decay mass $M$ of the plaquette correlation function in dependence on $\bar\beta^{-1/2}$

$$\eta(t) = \sum_{x_1} < \Phi_{(x_1,t)}\Phi_{(0,0)} >, \quad \Phi_x = P_x - P_x^{-1}, \quad (6)$$

$P_x$: Wilson Plaquette, see Figure 1. Perfect scaling appears for $\bar\beta=10$. A similar investigation for the heavy quark potential can be found in [9].

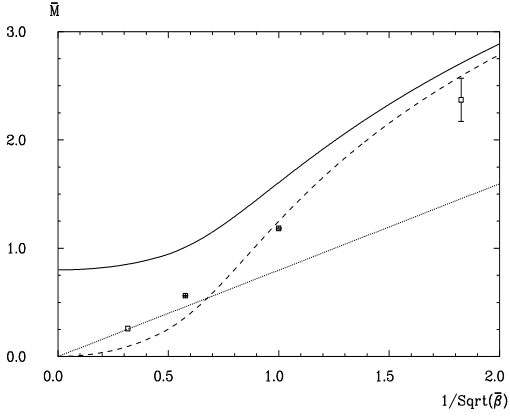

Figure 1. The gauge field mass vs. $\bar\beta^{-1/2}$. Dotted line: Scaling, full and dashed line: 0. and 1. order of an unquenched strong coupling expansion.

## 3. LATTICE FERMION DENSITIES

In order to study the structure of fermion observables, we relate lattice bilinears with the continuum densities. For that we use the relation of irreducible representations (irreps) of continuum and lattice symmetry in the following table:

Relation of continuum and lattice irreps

| cont. bilin. | $(I_R, Q_R; I_L, Q_L; l^{(\pi)})$ | $(I_R, I_L, Q_V, l^{(\pi)})$ | lattice bilinears |
|---|---|---|---|
| $j_1^0$ | $(0,0;$ | $(0,0,$ | $M^{1+}_{(0,0)}$ |
| $j_2^0$ | $0,0;2)$ | $0;2)$ | $M^{2+}_{(0,0)}$ |
| $j_1^3$ | | | $M^{2-}_{(1,1)}$ |
| $j_2^3$ | | | $M^{1-}_{(1,1)}$ |
| $j_1^1$ | $(1,0;$ | $(1,0,$ | $M^{0}_{(0,1)}$ |
| $j_2^2$ | $0,0;2)$ | $0;2)$ | $M^{0}_{(1,0)}$ |
| $j_2^1$ | | | $M^{12}_{(0,1)}$ |
| $j_1^2$ | | | $M^{12}_{(1,0)}$ |
| $s^0$ | | | $M^{0}_{(0,0)}$ |
| $s_5^3$ | | $(\frac{1}{2},\frac{1}{2},$ | $M^{0}_{(1,1)}$ |
| $s_5^1$ | | $0;0^+)$ | $M^{2-}_{(0,1)}$ |
| $s_5^2$ | $(\frac{1}{2},1;$ | | $M^{1-}_{(1,0)}$ |
| $s_5^0$ | $\frac{1}{2},-1;0)$ | | $M^{12}_{(0,0)}$ |
| $s^3$ | | $(\frac{1}{2},\frac{1}{2},$ | $M^{12}_{(1,1)}$ |
| $s^1$ | | $0;0^-)$ | $M^{1+}_{(0,1)}$ |
| $s^2$ | | | $M^{2+}_{(1,0)}$ |

The continuum densities (1.col.) are in the irreps of the continuum group (2.col.), belonging to ISC ($j_\mu^0$), IVC ($j_\mu^i$) and (pseudo)scalars ($s^i_{(5)}$). These are given by right-handed and left-handed isospin and charge and by spin and parity. Under the anomaly the scalar octet splits (3.col.). These continuum irreps split again into 2-dim. lattice irreps. We give them by their realizations by minimal lattice bilinears

$$M_L^0(x) = e^{i\pi(L,x)} \bar\chi_x \chi_x, \qquad (7)$$

$$M_L^{\mu +}(x) = (1/2)e^{i\pi(L',x)}[\bar\chi_x U_{[x,\mu]} \chi_{x+e_\mu} + h.c.],$$

$$M_L^{\mu -}(x) = (i/2)e^{i\pi(L',x)}[\bar\chi_x U_{[x,\mu]} \chi_{x+e_\mu} - h.c.],$$

$$M_L^{12}(x) = \frac{1}{8}e^{i\pi(L',x)} \sum_{(\mu,\nu)} \bar\chi_x U_{[x,\mu]} U_{[x+e_\mu,\nu]} \chi_{x+e_{\mu\nu}}.$$

$L=(L_1, L_2)$, $L_\mu=0,1$ is the flavour quantum number of these fields. For $M_L^{2\pm}$, $M_L^{12}$, $L'$ is given by $(L'_1, L'_2)=(1-L_1, L_2)$, for $M_L^{1\pm}(x)$ $L'=L$. In this way the lattice bilinears are related to the continuum bilinears. For an explicit treatment of these symmetry properties we refer to [1].

For some of these fields we evaluated the correlation functions with fixed spatial momenta $p = 0, \pi/3$ from our MC ensembles [1] ($x = (x_1, t)$)

$$m_{L_1}^\alpha(t;p) = \sum_{x_1} e^{ipx_1} <M_{(L_1,0)}^\alpha(x)M_{(L_1,0)}^\alpha(0)>, \quad (8)$$

We separated the time dependence by

$$m_{L_1}^\alpha(t;p) = \sum_i c_i(\epsilon_i)^t \cosh[E_i(p)(T/2-t)]. \quad (9)$$

For each energy $E_i$ in this fit the quantum number $L_2$ is given by the sign $\epsilon_i$. The separation of these 2 cases in the fit means the separation of quantum numbers from the current and scalar sector.

In the current sector we found only one energy, respectively, like in the continuum. The following table gives the continuum result (C), $\bar\beta=10$ results with c-contributions only, $Q_{top}=0$ (a), and c- and d-contributions at, $Q_{top}=0,\pm1$ (b).

Fit parameters for IVC and ISC case

| IVC | $m_+^0(t;\pi/3)$ | | $m_-^{1-}(t;\pi/3)$ | |
|---|---|---|---|---|
| | $E_1^-$ | $-C_1^-$ | $E_1^-$ | $C_1^-$ |
| C | 1.0472 | 1/3 | 1.0472 | 1/3 |
| a | .9409(4) | .2662(2) | .9429(5) | .2595(4) |
| b | .9409(5) | .2663(3) | .9429(6) | .2596(5) |
| ISC | $m_+^{1+}(t;0)$ | | $m_+^{1+}(t;\pi/3)$ | |
| | $E_1^+$ | $-C_1^+$ | $E_1^+$ | $-C_1^+$ |
| C | 0.252 | — | 1.0772 | — |
| a | .15(25) | .001(1) | .9429(5) | .2595(4) |
| b | .267(5) | .095(2) | .972(1) | .2663(5) |

For the IVC we found energies belonging to mass zero. For $p=0$ a IVC contribution is not significant. For $p=\pi/3$ only the connected contribution with $Q_{top}=0$ is important. The correlation functions of two different lattice fields became similar, indicating flavour restoration.

The c-contributions for the ISC are equal to those of the IVC. The d-contributions shift the mass to the value expected from the continuum. This effect is related to the axial anomaly.

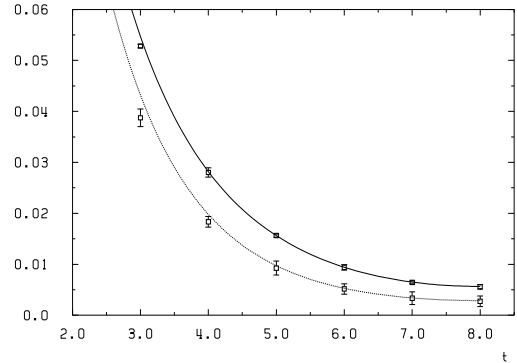

Figure 2. The correlation functions of 1-link fields at $\bar\beta = 10$, compared with the continuum result.

In the scalar sector we compared the data directly with the continuum curves. The values for the local field fit the continuum result. We considered 2 1-link fields belonging to $s_+$ and $s_-$ in the continuum. The corresponding continuum correlation functions are splited by the ZM contributions, indicating anomalous axial symmetry breaking. We could reproduce this on the lattice, see Figure 2 (we multiplied both lattice 1-link correlations with a renormalization factor 1.15). The splitting on the lattice is due to the $Q_{top} = \pm1$ contributions. They are large and thus balance the small fermion determinant in these sectors.